# Comment on "Time Step Sensitivity of Nonlinear Atmospheric Models: Numerical Convergence, Truncation Error Growth, and Ensemble Design" Teixeira et al. (2007)

Lun-Shin Yao[a]

Mechanical and Aerospace Engineering, Arizona State University, Tempe, Arizona

Dan Hughes

Hughes and Associates, Porter Corners, New York

Over the last one-half century due to continuous improvement in computers, numerical simulations have been developed into important tools, if not the only ones, to solve differential equations related to real problems from atomic to astronomic scales. Consequently, the number of engineers and scientists engaged in the use of such tools has increased dramatically. Computation has become, in many ways, the modern version of "mathematical analysis," and the kinds of problems analyzed by numerical methods keeps on growing. Most of the truly challenging problems, such as those related to chaos or turbulence, are routinely approached in this way. Researchers and other practitioners confidently believe that their particular problems are actually being solved by merely running their computers. Unfortunately, the importance of careful checks of convergence, an expected outgrowth from "numerical analysis," are often ignored.

The sensitivity of computed results for chaos or turbulence to the size of integration time steps is not completely unknown in the numerical-analysis community. However, most people, who recognize the existence of this problem, optimistically believe that the statistical properties of such computed results are relevant to their problem. The apparent reason for this belief is the fact that so many individuals have done similar types of computations and have published vast numbers of papers and reports. How could it be that something is wrong, given all of this effort? Simply stated, it has become common practice to *not* check computational results since they have been accepted for such a long time. This situation may be partially due to the fact that the problem of error analysis for nonlinear differential equations has not been systematically studied. Under this family of circumstances, raising questions about the validity of numerical simulations of chaos or turbulence is viewed by some as an attack on their integrity.

---

[a]Address for correspondence: Mechanical and Aerospace Engineering, Arizona State University, Tempe, Arizona, 85287-6106.  E-mail: 2003.yao@asu.edu

We want to congratulate the members of the meteorology group at the Naval Research Laboratory in Monterey, California supported by the Office of Naval Research for publishing the important paper on their systematic study of time-step sensitivities for nonlinear differential equations of chaos and turbulence. Their firm conclusion is that computational chaos results of differential equations, which are sensitive to time steps, are simply errors, as are their ensemble averages. Their work is a landmark contribution, and will lead a long line of future studies, which will improve understanding of the amplification of truncation errors in computational chaos and turbulence.

Even though it does not alter the conclusion of their paper, an objection about their assumption that the result of the smallest time step is closest to the correct one can be made. The amplification mechanism for truncation errors has been analyzed from a geometric point of view for the Lorenz system [Yao 2005]. Two types of error amplifications, exponential and "explosive," were identified.

The more serious one is the explosive amplification that occurs when a trajectory penetrates a local separatrix near the z-axis for $z < 12$, thereby violating the differential equations. The local separatrix is the inset of the node located at the origin. It has been shown that the occurrence of a trajectory that penetrates this local separatrix is not a monotonic function of the integration time steps. Consequently, the smallest time step cannot be viewed as the closest "solution" to the correct one. In fact, no one knows where the correct trajectory is!

Our second point is that the explosive error amplification occurs repeatedly and randomly so it does not follow an exponential rule. Its cumulative effect might be correlated by an exponential function, but its properties are far more complex. An interesting open question is: Is there any relation between "error explosion" and "homoclinic explosion" of computed Lorenz attractors?

REFERENCES:

Teixeira, J., C. A. Reynolds, and K. Judd, 2007: Time step sensitivity of nonlinear atmospheric models: numerical convergence, truncation error growth, and ensemble design. *J. Atmos. Sci.*, **64**, 175-189.

Yao, L. S., 2005: Computed Chaos or Numerical Errors. Online Manuscript, http://arxiv.org/abs/nlin.CD/0506045.